# Response to Comment on "Perceptual Learning Incepted by Decoded fMRI Neurofeedback Without Stimulus Presentation";

# —How can a decoded neurofeedback method (DecNef) lead to successful reinforcement and visual perceptual learning?—

Kazuhisa Shibata[1,2], Yuka Sasaki[2,3], Takeo Watanabe[2,3,*], Mitsuo Kawato[2]

1 Department of Psychology, Graduate School of Environmental Studies, Nagoya University, Furo-cho, Chikusa-ku, Nagoya-shi, Aichi 464-8601, Japan

2 ATR Brain Information Communication Research Laboratory Group, 2-2-2 Hikaridai, Keihanna Science City, Kyoto 619-0288, Japan

3 Department of Cognitive, Linguistic, & Psychological Sciences, Brown University, 190 Thayer Street, Providence, RI 02906, USA

* To whom correspondence should be addressed to takeo_watanabe@brown.edu

**Abstract**

Huang (arXiv:1612.03270) argues that the perceptual learning induced by our decoded neurofeedback method (DecNef) can be explained by Hebbian synaptic plasticity of connections between V1/V2 and V3/V4 rather than that within V1/V2, and that reinforcement learning at a cellular level should not be possible, and thus challenges our conclusions. In this reply, first, we show that Huang's model is not compatible with our data which strongly suggest plasticity within V1/V2. Second, the results of our new analysis show that spontaneous activity is not random but largely accounted for by low-dimensional components. These and other results indicate that learning to induce a stimulus-associated template voxel pattern in V1/V2 within as few as 100 trials is likely to be accomplished by hyper-resolution and sub-voxel level control by DecNef.

Dr. Huang has raised four thoughtful concerns (*1*) regarding the interpretations and results of our study that demonstrated a successful application of a new decoded neurofeedback method (hereafter we call this method DecNef). Here, we will respond to each of these concerns. As pointed out in our paper (*2*), DecNef causes two types of learning. For the first type, subjects learn to induce a stimulus-associated template voxel pattern without the presentation of an actual stimulus. For the second type, perceptual learning occurs as a result of repetitive inductions of the stimulus-associated template pattern. Dr. Huang's first two concerns have much to do with the second type of learning, while the other two concerns are mainly concerned with the first type of learning. Here, we will first dispute each of these concerns.

**Concerns with changes in V1/V2 associated with perceptual learning**

(1) First, Dr. Huang indicates that the visual performance improvements found in our study can be attributed to a Hebbian learning model which assumes connectivity changes between V1/V2 and higher visual areas including V3/V4, rather than plasticity within V1/V2. In this model, learning (i.e., trial-by-trial update of weights between V1/V2 and V3/V4) occurs when a specific input pattern (activation pattern of V1/V2) and a corresponding output pattern (activation pattern of V3/V4) are co-activated.

If this point was indeed true, fMRI activation patterns in V3/V4 should have changed in concert with fMRI activation patterns in V1/V2 on a trial basis. However, our results showed no such tendency (see Fig. S9 in *Shibata et al. (2)*). Thus, Huang's hypothesis that learning results from connectivity changes between V1/V2 and V3/V4 or higher areas is not consistent with the results of our study, while it is possible that perceptual learning in our study resulted from connectivity changes within V1/V2 circuits (*3*).

(2) Dr. Huang suggests that the mere induction of a stimulus-associated template voxel pattern in V1/V2 that was obtained in the decoder construction stage that was conducted prior to the induction stage may not cause perceptual learning.

Therefore, higher areas in the visual hierarchy than V1/V2 should be involved in perceptual learning.

This point does not consider the fact that repeated synchronous activity of different units can strengthen the connectivity between those units while the activity of the units themselves remains unchanged. Such connectivity changes have been observed in many types of neural activities including associative long-term potentiation (LTP) (*4-6*). Thus, it may well be rationalized that repeated inductions of a stimulus-associated template voxel pattern in V1/V2 changes neural connectivity within V1/V2 without changing the voxel patterns themselves.

In addition, a high correlation between the target-orientation likelihoods in V1/V2 in the induction stage and sensitivity changes for the target orientation in the testing stage (Fig. 3E in *Shibata et al.* (*2*)) provides strong evidence that plasticity within V1/V2 did occur in association with perceptual learning.

Thus, from these theoretical and empirical viewpoints, it is highly unlikely that the perceptual learning shown in our study resulted from changes in areas higher than V1/V2.

**Concerns with learning of a stimulus-associated template pattern**

(3) Huang's third point concerns the "curse of dimensionality". The dimensional space of potential voxel activity patterns is very large. If each voxel pattern is randomly searched in a standard reinforcement learning paradigm, it would be impossible, argues Huang, for the visual system to find an efficient solution within as few as 100 trials because the search space is too high-dimensional. Our decoding algorithm, called "Sparse Logistic Regression (SLR)" (*7, 8*), automatically selects 239±29 voxels of V1/V2 (see Supporting Online Material (SOM) of *Shibata et al*. (*2*)), and subjects either increase or decrease different subsets of these voxels (Fig. S8 of *Shibata et al*. (*2*)) in the induction stage. For each voxel, there are at least two possible actions: increasing or decreasing activity. When 239 voxels are taken into account, there exist $2^{239}$ (more than 8.8 x $10^{71}$) possible combinations of actions. If these actions are tested serially, as Huang postulated, even the projected lifespan of the universe would be too short a time for the visual system to test them all, let alone 100 trials in a day.

Many previous studies that examined spontaneous activities of the visual cortex in animals have indicated that the above point by Huang is not true. It has been found that spontaneous activities are by no means random but contained in the low-dimensional manifold that is spanned by neural responses to natural scenes or visual images, and are autonomously generated by connections between or within columns (*9-12*). Thus, spontaneous activities at sub-voxel levels must be severely constrained by existing connections, and actually contain neural response signals related to natural scenes including the visual feature information such as orientation (*11*). These low-dimensional sub-voxel spontaneous activities should induce similar low-dimensional voxel spontaneous activities during the induction stage because the mapping from the neuronal to voxel levels is unique and well defined.

We have four lines of strong empirical evidence that support this reasoning. One line of evidence is that the subjects in *Shibata et al*. (*2*) could actually learn to induce a stimulus-associated template voxel pattern. The occurrence of such learning proves that the search space is low-dimensional. That is, the spontaneous activation in the induction stage was indeed of only low dimension at the voxel level.

The second line of evidence is provided by the results of principal component analysis (PCA) that has been newly applied to fMRI data from both the decoder construction and induction stages (see Methods for details). While the analysis utilizing the SLR decoder in *Shibata et al*. (2) focused on the likelihood of the target orientation on only one-dimensional scale, the PCA here is model-free and adopts multiple-dimensional scales to represent principal components (PCs) on a fair ground. Results of the PCA indicate that as large as 60% of the variance in these data was accounted for (60% variance accounted for (VAF)) by only top 10 principal components (PCs) both in the decoder construction stage (blue in Fig. 1A) and in the induction stage (cyan in Fig. 1A). If the spontaneous activity were indeed purely random as assumed by Huang, then as many as 239 PCs, which correspond to the mean number of voxels across the subjects that SLR took into account (see Methods), should be necessary to account for the individual fMRI data. In that case, only around 4.2 % (10 dimensions out of 239 dimensions = ~0.042) VAF at maximum would be able to account for by 10 PCs, in contrast to the 60% VAF in our results. Furthermore, in our analysis, the top 10 PCs obtained from the decoder construction stage could then account for more than 40% of the variance in the induction stage (magenta in Fig. 1A). That is, more than 40% of the spontaneous voxel activity during the induction stage can be accounted for by orientation-stimulus driven activities. Thus, these results indicate low dimensionality in both of the orientation-stimulus driven activities during the decoder construction stage and the spontaneous voxel activities during the induction stage. Moreover, the major portion of the variance in the spontaneous voxel activity during the induction stage was accounted for by stimulus-driven activities. That is, the stimulus-driven activities during the induction stage were close to spontaneous activities without stimulus presentation.

The third line of evidence was obtained through comparisons between activities during the decoder-construction stage and activities during the induction stage. If the fluctuations of spontaneous activities of each voxel in the induction stage were actually uncorrelated with those of neighboring voxels, as assumed by Huang, then the distributions of spontaneous activities in the axes representing orientation in the PCA analysis should be uniform and narrow. We examined the validity of this prediction by means of PCA applied to individual trial data of a representative subject in the decoder-construction stage (Fig. 1C). We sought and found the three PCs that best represent orientation information. Fig. 1C is from the decoder construction stage, and plots each trial as a dot, along the three dimensional axes consisting of the third, ninth and tenth PCs of the data. In each trial, one orientation out of three was presented. The color (red, blue, and green) of dots represents one of three different stimulus orientations presented with the subject. The position of each dot represents its PC weights. Each of the three colored ellipsoids represents the distribution of the dots in the same color. The mean and the standard deviations are represented as the location of the center and the three radii of the ellipsoid, respectively. As is shown in Fig. 1C, each of colored ellipsoids tends to fall into three distinct clusters (see also Supporting Movie). This indicates that orientation information was well represented by the selected 3 PCs. Fig. 1C also shows that the brain activations in response to the presentation of an identical orientation in the decoder construction stage fluctuate with a certain variation but that they are mostly constrained by intrinsic connections at a sub-voxel (probably neuronal) level (*10*).

The fourth line of evidence indicates that even the spontaneous voxel activities for the first 30 trials of day 1 of the induction stage, during which no learning occurred (SOM of *Shibata et al.* (*2*)), are constrained by existing neural connections that reflect orientation information. We plotted data from the first 30 trials of day 1 (Fig. 1D) along the same axes determined in the analysis of the decoder construction stage (Fig. 1C). In Fig. 1D, each color represents the classified orientation by the SLR decoder. The large open ellipsoid in Fig. 1D was computed in the same way as in Fig. 1C. Note that the locations, shapes and sizes of the ellipsoids in Figs. 1C and D are very similar to each other. This indicates that the distribution of the spontaneous voxel activities even for the first 30 trials on the first day of the induction stage (Fig. 1D), for which no learning was observed, is very similar to the distribution of stimulus-driven activations (Fig. 1C). If the assumption of uncorrelated fluctuations of each voxel activity was actually correct, as Huang assumes, then the ellipsoid in Fig. 1D should be a complete sphere with the radius of 1. However, the shape of the ellipsoid obtained from the true data in Fig. 1D is highly different from that of the sphere based on Huang's assumption. In addition, the volume of this erroneously predicted sphere is as small as 1/14 of the ellipsoid based on our data in Fig. 1D. Thus, the results of our analysis strongly suggest that the constrained spontaneous voxel activities are generated by neural connections at sub-voxel levels, which had been acquired through one's visual experiences including orientation information before learning in the induction stage occurred (*11*).

All these results demonstrate that spontaneous activities at the voxel level are far from being random and strongly constrained by existing neural connections for representing visual information. This indicates that the assumption by Huang is incorrect. The results of these analyses are in accord with the model that voxel activity patterns represent visual information that already exists even at the beginning of the induction stage due to intrinsic connections as shown in Fig. 1D and these patterns were selectively reinforced by DecNef later in the induction stage. Therefore, the curse of dimensionality is invalid in the DecNef procedure.

(4) Huang's last concern notes that even if activities at the fMRI voxel level are controlled by DecNef, neural activities at sub-voxel levels, including the columnar, neural and synapse levels, may be beyond control. fMRI measurement is, in this sense, coarse, and therefore many states at sub-voxel levels may correspond to one state at the fMRI voxel level. Because of this "many to one mapping" issue, Huang argues that it should not be possible to conclude that DecNef causes plasticity at a cellular level and perceptual learning associated with changes in V1/V2.

In the discussion to reply to the third concern, we demonstrated that voxel spontaneous activities are constrained by neural connections that reflect visual information and are therefore low-dimensional. Thus, the curse of dimensionality is not valid in the DecNef procedure. Here we will prove that DecNef is not only constrained by but also controls neural connection.

First, the occurrence of perceptual learning in our study (*2*) implies that DecNef can control neural-level networks, since perceptual learning is regarded as a manifestation of neural plasticity.

Second, since a visual system's orientation-decoding algorithm can achieve much higher resolution than each voxel size, as has been demonstrated by Kamitani and Tong (*13*), DecNef should generate and induce corresponding hyper-resolution signals, since DecNef contains an orientation decoding algorithm. Thus, the fact that perceptual learning occurred as a result of repeated induction of such hyper-resolution signals in our study (*2*) indicates that DecNef can control hyper-resolution signals that should be at a sub-voxel (probably neural) level. As discussed in the previous section, perhaps the DecNef controls such sub-voxel signals by selectively reinforcing different, but largely constrained, activity patterns that are classified by the decoding algorithm in DecNef.

Third, the result that VAF increased across days during the induction stage (Fig. 1B) supports the idea that DecNef can successfully control sub-voxel, perhaps neuronal, activity at an orientation representation level. This indicates that as learning proceeds, spontaneous voxel activities during the induction stage becomes increasingly similar to stimulus-driven activities during the decoder construction stage.

Fourth, an analysis of distribution of the spontaneous activity in association with perceptual learning indicates that DecNef produced changes at an orientation representation level. We further tested whether the learning by DecNef in our study (*2*) occurred only with respect to the likelihood output of the decoder without actually committing activities related to orientation representation at neural levels. The results indicate that most of the spontaneous activities on the fifth day (Fig. 1E) of the induction stage were classified as the target orientation. In addition, the spontaneous activities were very similar to that of the target stimulus-driven activities (red dots in Fig. 1C) (see the similarities in positions, sizes and shapes of the two red ellipsoids in Figs. 1C and E). Thus, DecNef not only causes successful learning with respect to the likelihood output of the decoder, but also produced voxel activity patterns that are very similar to those induced by the actual target visual stimuli in the coordinate frame relevant to orientation information.

All the result of our analyses are in accord with the model that DecNef can successfully control orientation representations which should be at a sub-voxel, perhaps neural, level.

In summary, Huang raised counterarguments concerning (i) changes in V1/V2 associated with perceptual learning and (ii) the plausibility of learning a stimulus-associated template pattern via DecNef protocols. Both empirical evidence and logical considerations indicate that our conclusions in *Shibata et al.* (2) are valid. Results of our new analysis indicate that DecNef represents and controls higher resolution signals than those at the voxel level.

**Figures**

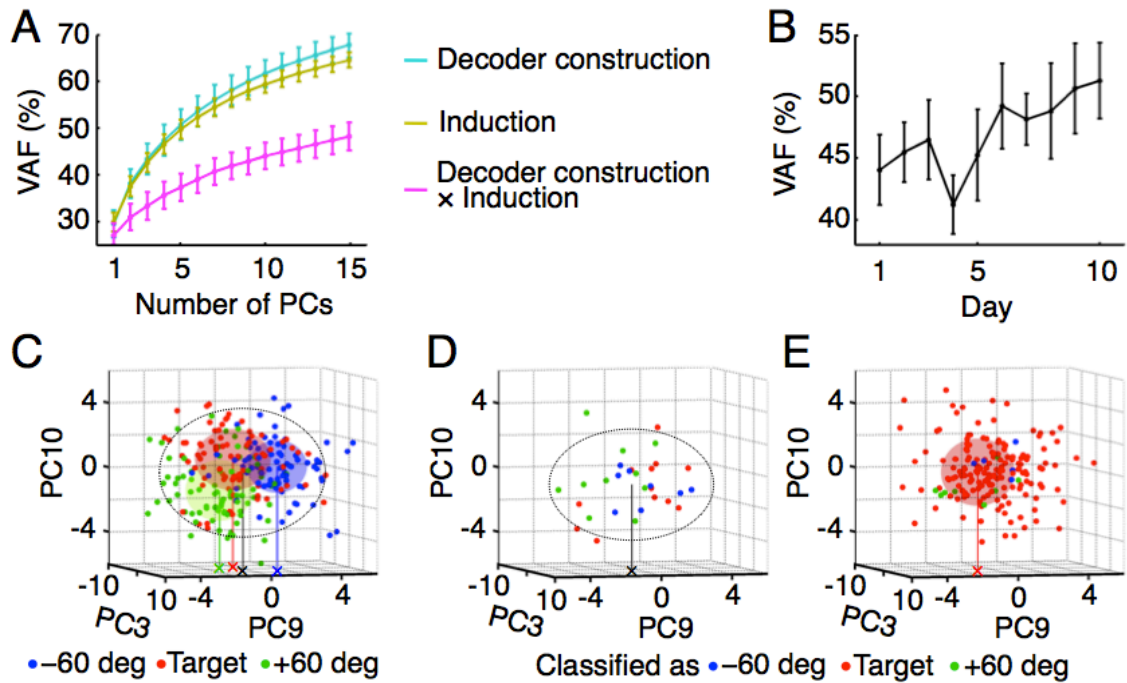

Fig. 1. Results of PCA. (A) Mean (±SE) variance accounted for (VAF) as a function of the number *(n)* of principal components (PCs) in the decoder construction stage (cyan), the induction stage (yellow), and for the induction-stage data fitted by the top *n* PCs obtained for the decoder construction stage (magenta, decoder construction × induction). (B) Mean (±SE) time-course of VAF as a function of the training day in the induction stage. VAF represents how much variance of the voxel activity within each day of the induction stage was accounted for by the top 10 PCs in the decoder construction stage. (C) Plot of data in the decoder construction stage from one representative subject within the axes consisting of 3 major PCs. Each colored dot represents one trial in which one of the 3 orientation stimuli was presented with the subject. The three PCs were selected so that orientation information is best decoded in this framework. Each colored ellipsoid represents the mean±SD range of a distribution of the trials for each orientation stimulus. The large open ellipsoid represents the distribution of all data points. The center and three radii of the ellipsoid represent the mean and double lengths of three standard deviations along three axes, respectively. Each colored vertical line represents the

**projection of the center of an ellipsoid on the PC3-PC9 plane. (D, E) Data from the induction stage for the day 1 and day 5 for the same subject as in C were plotted using the same 3 PC axes of the decoder construction stage as in C. The color of each dot is determined from orientation classification by the sparse logistic regression decoder of that data. The broken line ellipsoid in D and the red ellipsoid in E were computed in the same manners as in C.**

## Methods

Spatial principal component analysis (PCA) was applied to voxel activities of V1/V2 in the decoder construction and induction stages (see *Shibata et al.* (*2*) for details for V1/V2 specification and voxel activity calculation). The mean (±SE) number of voxels selected by the sparse logistic regression decoder (see *Shibata et al.* (*2*) for details) was 239±29 in V1/V2 across 10 subjects. The selected voxels were used for the PCA. The number of the utilized voxels corresponds to the full dimensions of PCA.

In the decoder construction stage, each of 3 orientation stimuli was presented 80 times see *Shibata et al. (2)*. Thus, there were 239 (in average across the subjects) voxel patterns for each subject. PCA calculates a transformation matrix to project these voxel patterns to principal components (PCs). The number of PCs is the same as the number of utilized voxels in V1/V2. We calculated the variance of each PC and sorted the PCs according to the variance. The "variance accounted for" (VAF) for top *n* PCs was defined by the summation of the variance of top *n* PCs divided by the total variance of all PCs. VAF for *n* PCs indicates how much proportion of voxels patterns are accounted for by *n* PCs.

In the induction stage, the subjects conducted at most 180 trials for each day (*Shibata et al. (2)*). The mean (±SE) number of trials for the subjects who participated in the 5-day induction stage was 788±38, and the mean (±SE) number of trials for the subjects participated in the 10-day induction stage was 1638±44. As in the decoder construction stage, we calculated the transformation matrix and the PCs using all trials. VAF was calculated for each day using the same method in the decoder construction stage.

In addition, we applied the transformation matrix calculated in the decoder construction stage for the voxel patterns in the induction stage to quantify how similar a structure of voxel patterns in the decoder construction stage is with a structure of voxel patterns in the induction stage. VAF was calculated for each day by the same method in the decoder construction stage.